\documentclass[onecolumn,prd,showpacs]{revtex4}
\usepackage{graphicx}
\usepackage{amssymb}
\begin{document}

\title{\bf A Unified Scheme of Central Symmetric Shape-Invariant Potentials}
\author{T. Koohrokhi}
\email{t.koohrokhi@gu.ac.ir}

\author{A. Izadpanah}
\email{a.izadpanah@gu.ac.ir}

\author{M. Gerayloo}

\affiliation{Department of Physics, Faculty of Sciences, Golestan University\\ Gorgan, Iran}

\begin{abstract}
Most physical systems, whether classical or quantum mechanical, exhibit spherical symmetry. Angular momentum, denoted as $\ell$, is a conserved quantity that appears in the centrifugal potential when a particle moves under the influence of a central force. This study introduces a formalism in which $\ell$ plays a unifying role, consolidating solvable central potentials into a superpotential. This framework illustrates that the Coulomb potential emerges as a direct consequence of a homogenous ($r$-independent) isotropic superpotential. Conversely, a $\ell$-independent central superpotential results in the 3-Dimensional Harmonic Oscillator (3-DHO) potential. Moreover, a local $\ell$-dependent central superpotential generates potentials applicable to finite-range interactions such as molecular or nucleonic systems. Additionally, we discuss generalizations to arbitrary $D$ dimensions and investigate the properties of the superpotential to determine when supersymmetry is broken or unbroken. This scheme also explains that the free particle wave function in three dimensions is obtained from spontaneous breakdown of supersymmetry and clarifies how a positive 3-DHO potential, as an upside-down potential, can have a negative energy spectrum. We also present complex isospectral deformations of the central superpotential and superpartners, which can have interesting applications for open systems in dynamic equilibrium. Finally, as a practical application, we apply this formalism to specify a new effective potential for the deuteron.
\end{abstract}

\maketitle
\section{Introduction}
Unified schemes in physics represent a fundamental pursuit to distill the complexity of natural phenomena into cohesive frameworks, fostering a deeper understanding of the underlying principles governing the universe. These schemes aim to unify disparate theories and forces, transcending apparent divisions. A striking example is the Standard Model of particle physics, which successfully unifies electromagnetic, weak, and strong nuclear forces, providing a comprehensive description of the fundamental particles and their interactions \cite{Goldberg2017}. Supersymmetry is a theoretical extension of the Standard Model can resolve some important problems by unifying the description of forces with that of matter \cite{Ramond1971,Neveu1971,Gel1971,Volkov1973,Wess1974}. In the context of quantum mechanics (QM), supersymmetric quantum mechanics (SUSY-QM) may be considered as modernized version of Darboux transformation explains how two supersymmetric partner potentials (superpartners) unify to a single superpotential \cite{WITTEN1981513,Cooper1995,Gangopadhyaya2017,Axel2021}.

Trying to obtain the exact solutions has been an issue of great interest since origin of QM and helps to better understanding, classifying, generalizing and using quantum mechanical systems as well as to test some approximation methods. Exactly solvable potentials (ESPs) referred to the potentials that all the eigenvalues, finite or infinite in number, and the corresponding eigenfunctions can be obtained explicitly \cite{ginocchio1984class,spiridonov1992exactly,dutt1988supersymmetry}. The exploring has been categorized analytically solvable problems in QM into exactly \cite{levai1992some}, conditionally exact \cite{Dutta2010,nigmatullin2022unification}, quasi-exact \cite{Finkel1996,tkachuk1998quasi,gangopadhyaya1995methods} and next-to-exact \cite{Znojil1997,znojil1994classification} or semi-exactly  \cite{Dong2016,kar2016,karwowski2014biconfluent} solvable groups.

By using SUSY-QM formalization, it has been demonstrated that shape-invariant potentials (SIPs) are exactly solvable \cite{chuan1991exactly,acar2023unusual}. In addition of their exact solvability, SIPs utilize to test of exactness of approximation methods, e.g., the supersymmetric WKB (SWKB) approximation \cite{Gangopadhyaya2021,GANGOPADHYAYA2020126722}. By definition, SIPs have the same dependence on main variables and differ from each other in the value of parameters. For these types of potentials, there are several kinds of classifications based on their properties, transformation mechanisms, explicit dependence on $\hbar$, as well as their bound-state wave functions. From the point of view of transformation mechanism, SIPs is assort into three main categories: translational or additive \cite{Dutt1988}, multiplicative or scaling \cite{Spiridonov1992}, and cyclic \cite{SUKHATME1997401} SIPs.

In one-dimensional quantum mechanical problems, additive SIPs that are generated by $\hbar$-independent additive shape-invariant superpotentials (SISs) are most well-known and commonly used so that they are called "conventional" \cite{MALLOW2020126129}. It is proven that the conventional SISs form a complete set and their bound-state wave functions are expressed in terms of classical orthogonal polynomials. On the other hand, some rational extensions or isospectral deformations of SIPs led to the discovery of exceptional orthogonal polynomials \cite{Assi2021,sym12111853}. Furthermore, in two and three dimensions (3$D$), new classes of exactly solvable non-central potentials can be produced by rational extensions of corresponding central potentials (CPs) \cite{Kumari2018}. In this respect, we present here 3$D$ central additive SIPs in the framework of a unified scheme. A central superpotential (CS) in the present study generates ESPs under various conditions, so the formalism can be generalized to any arbitrary $D$-dimension straightforwardly.

A significant recent advancement in QM involves the expansion of SIPs into the complex domain \cite{Kooh2023}. These newly introduced generalized non-Hermitian SIPs share an isospectral relationship with their Hermitian counterparts. In conventional QM, Hermitian Hamiltonians are exclusively employed as descriptors for closed systems. Although the Hermiticity condition ensures that the eigenvalues of the energy are real, it is, in addition to being restrictive, merely a mathematical condition and does not offer a profound understanding of quantum systems. On the other hand, both experimentally and theoretically, it has been demonstrated that the energy eigenvalues of non-Hermitian Hamiltonians, associated with open systems in dynamic equilibrium with their surroundings, can also take real values \cite{Ben18}. Consequently, we have also investigated complex isospectral deformations of central SIPs, anticipating interesting theoretical and practical applications in quantum systems.

The rest of the paper is organized as follows: In Sec. II, we provide a brief introduction to SUSY-QM and its application to 3D central symmetry. Section III delves into several special cases of the CS, discussing the methodology for deriving well-known CPs from it. The aspects of broken and unbroken of SUSY and their relevance to SIPs are explored in Sec. IV. In Sec. V, we highlight how the CPs can be extended to  arbitrary $D$-dimensions, and then the hierarchy of the central P\"{o}schl-Teller potential is explored in Sec. VI. Section VII focuses on obtaining the complex isospectral deformations of CPs by generalizing the superpotential to the complex domain. As a practical application and illustration, Sec. VIII examines the deuteron using an effective potential. Ultimately, in Sec. IX, we summarize our findings and provide some directions for future research.

%****************************************** Section 2: SUSY-QM for 3-Dimensions in Central Symmetry ************************************
 \section{SUSY-QM for 3-Dimensions in Central Symmetry}
In spherical coordinates, a CP depends solely on the radial variable $r$. A formal procedure involves separating the radial and angular variables from the wave function, expressed as $\psi_{n\ell m}(r,\theta, \phi)=\Re_{n\ell}(r)Y_{\ell m}(\theta, \phi)$, and introducing a change of variable $\Re_{n\ell}(r)=r^{-1}u_{n\ell}(r)$. Consequently, the radial term of the time-independent Schr\"{o}dinger equation (SE) can be summarized as follows \cite{Gangopadhyaya2017}.
\begin{equation}
\frac{\hbar^{2}}{2m}\frac{u^{\prime\prime}_{0\ell}(r)}{u_{0\ell}(r)}=V_{1}(r,\ell),
\end{equation}
where the subscript "0" denotes the ground state (GS), indicating the corresponding potential,
\begin{equation}
V_{1}(r,\ell)=v(r,\ell)+V_{\rm Cef}(r,\ell),
\end{equation}
consists of the centrifugal potential,
\begin{equation}
V_{\rm Cef}(r,\ell)=\frac{\hbar^{2}}{2m}\frac{\ell(\ell+1)}{r^{2}},
\end{equation}
which always appears in spherical symmetry for non-zero angular momenta. Additionally, the potential,
\begin{equation}
v(r,\ell)=V(r)-E_{0\ell},
\end{equation}
includes the GS energy $E_{0\ell}$ and the CP, $V(r)$. Equation (1) implies that the potential $V_{1}(r,\ell)$ has a GS energy of zero, denoted as $E^{(1)}_{0\ell}=0$. This condition is a crucial requirement for unbroken SUSY. According to SUSY-QM, the superpotential is defined as the logarithmic derivative of the GS wave function,
\begin{equation}
W(r,\ell)=-\frac{\hbar}{\sqrt{2m}}\frac{d}{dr}\ln{u_{0\ell}(r)},
\end{equation}
The SE reduces to the nonlinear Riccati equation (RE). In light of this, the potentials $V_{1}$ and its superpartner $V_{2}$ satisfy the following relations,
\begin{equation}
\left \{\begin{array}{ll} V_{1}(r,\ell)=W^{2}(r,\ell)-\frac{\hbar}{\sqrt{2m}}W^{\prime}(r,\ell))\\
V_{2}(r,\ell)=W^{2}(r,\ell)+\frac{\hbar}{\sqrt{2m}}W^{\prime}(r,\ell), \end{array}
\right.
\end{equation}
Now, we propose a general ansatz for the corresponding superpotential as follows,
\begin{equation}
W(r,\ell)=\frac{\hbar}{\sqrt{2m}}\left\{w(r,\ell)+w_{\rm CefS}(r,\ell)\right\},
\end{equation}
Here, the first term, $w(r,\ell)$, generates the CP, and we refer to it as the "central superpotential". Moreover, the second term, $w_{\rm CefS}(r,\ell)=-\frac{\ell+1}{r}$, generates the centrifugal potential, and we term it the "centrifugal superpotential" (CefS). As we will discuss in the next section, the CS is a function determined by shape invariance in different situations.

By substituting the superpotential into RE, the superpartners are obtained as follows,
\begin{equation}
\left \{\begin{array}{ll}
V_{1}(r,\ell)=\frac{\hbar^{2}}{{2m}}\left\{w^{2}(r,\ell)-w^{\prime}(r,\ell)-2w(r,\ell)\frac{\ell+1}{r}+\frac{\ell\left(\ell+1\right)}{r^{2}}\right\}\\
V_{2}(r,\ell)=\frac{\hbar^{2}}{{2m}}\left\{w^{2}(r,\ell)+w^{\prime}(r,\ell)-2w(r,\ell)\frac{\ell+1}{r}+\frac{\left(\ell+1\right)\left(\ell+2\right)}{r^{2}}\right\}.
\end{array}
\right.
\end{equation}

In the context of SUSY-QM, a potential is considered shape-invariant if its superpartner has the same spatial dependence as the original potential, with possibly altered parameters. Therefore, the superpartners $V_{1}$ and $V_{2}$ are SIPs if the remainder $R_{\ell}$ is independent of $r$ \cite{Gomez2014}. By considering $\ell$ as a parameter, the remainder for additive SIPs is obtained as,
\begin{eqnarray}
R_{\ell}&=&V_{2}(r,\ell)-V_{1}(r,\ell+1)\nonumber\\
&&=\frac{\hbar^{2}}{{2m}}\Big\{w^{2}(r,\ell)-w^{2}(r,\ell+1)+w^{\prime}(r,\ell)+w^{\prime}(r,\ell+1)\nonumber\\
&&+2w(r,\ell+1)\frac{\ell+2}{r}-2w(r,\ell)\frac{\ell+1}{r}\Big\}.
\end{eqnarray}

In central symmetry, we are allowed to separate the dependencies of $r$ and $\ell$ in the CS, i.e., $w(r,\ell)\equiv f(r)g(\ell)$. If $g(\ell+1)$ is a function of $g(\ell)$, represented as $g(\ell+1)=G(\ell)g(\ell)$, then Eq. (9) transforms to,
\begin{eqnarray}
R_{\ell}&=&\frac{\hbar^{2}}{{2m}}\Big\{g^{2}(\ell)\left[f^{2}(r)-f^{2}(r)G^{2}(\ell)\right]
+g(\ell)\left[f^{\prime}(r)+f^{\prime}(r)G(\ell)\right]\nonumber\\
&&+2\frac{g(\ell)}{r}\left[f(r)G(\ell)(\ell+2)-f(r)(\ell+1)\right]\Big\}.
\end{eqnarray}
Factorizing this expression yields,
\begin{eqnarray}
R_{\ell}&=&\frac{\hbar^{2}}{{2m}}\Big\{w^{2}(r,\ell)\left[1-G^{2}(\ell)\right]+w^{\prime}(r,\ell)\left[1+G(\ell)\right]\nonumber\\
&&+2\frac{w(r,\ell)}{r}\left[G(\ell)(\ell+2)-(\ell+1)\right]\Big\}.
\end{eqnarray}
Ultimately, solving this first-order differential equation results in the general form of the CS for $G(\ell)>1$ and $R_{\ell}>0$,
\begin{equation}
w(r,\ell)=\left(\frac{B_{\ell}}{G(\ell)-1}\right)\frac{J(A_{\ell}+1,B_{\ell}r)+CY(A_{\ell}+1,B_{\ell}r)}
{J(A_{\ell},B_{\ell}r)+CY(A_{\ell},B_{\ell}r)},
\end{equation}
where $J$ and $Y$ are the Bessel functions of the first and second kinds, respectively, and $C$ is an integration constant. The coefficients $A_{\ell}$ and $B_{\ell}$ are given by,
\begin{equation}
\left \{\begin{array}{ll}
A_{\ell}=\left(\frac{G(\ell)-1}{G(\ell)+1}\right)\frac{2\ell+3}{2} \\
B_{\ell}=\sqrt{\left(\frac{G(\ell)-1}{G(\ell)+1}\right)\frac{2m R_{\ell}}{\hbar^{2}}}
\end{array}.
\right.
\end{equation}
In the following section, we will examine the remainder, Eq. (11), for different specific situations.

%******************************************Section 3 : The Special Cases ************************************
\section{The Special Cases}
%***************************************** Subsection 3.1 : G(\ell)=1 ************************************
\subsection{The $\ell$-Independent CS}
If $G(\ell)=1$, according to Eq. (11), the CS, $w(r)$, as well as the remainder, R, become independent of $\ell$, leading to,
\begin{equation}
R=\frac{\hbar^{2}}{{m}}\left\{w^{\prime}(r)+\frac{w(r)}{r}\right\}.
\end{equation}
Solving this first-order differential equation for CS gives,
\begin{equation}
w(r)=\frac{m}{2\hbar^{2}}Rr+\frac{C}{r}.
\end{equation}
For $R=2\hbar \omega$, substituting the CS into Eq. (8), we obtain the superpartners as,
\begin{equation}
\left \{\begin{array}{ll}
V_{1}(r,\ell)=\frac{1}{2}m\omega^{2}r^{2}+\frac{\hbar^{2}}{{2m}}\frac{C(C+1)+(\ell+1)(\ell-2C)}{r^{2}}-\hbar \omega\left(\ell+3/2-C\right)\\
V_{2}(r,\ell)=\frac{1}{2}m\omega^{2}r^{2}+\frac{\hbar^{2}}{{2m}}\frac{C(C+1)+(\ell+2)(\ell+1-2C)}{r^{2}}-\hbar \omega\left(\ell+1/2-C\right),
\end{array}
\right.
\end{equation}
In this context, the CP is given by,
\begin{equation}
V(r)=\frac{1}{2}m\omega^{2}r^{2},
\end{equation}
which represents the three-dimensional harmonic oscillator (3-DHO) potential.

%******************************** Subsection 3.2 : G(\ell)= -1 ************************************

\subsection{$G(\ell)= -1$}
On the other hand, for $G(\ell)=-1$, the CS, $w(r,\ell)$, is obtained from Eq. (11) as,
\begin{equation}
w(r,\ell)=-\frac{m}{\hbar^{2}(2\ell+3)}R_{\ell}r.
\end{equation}
Since $w(r,\ell)=f(r)g(\ell)$, assuming $f(r)=-\frac{m} {\hbar^{2}}r$ and $g(\ell)=\frac{R_{\ell}}{2\ell+3}$, the condition $G(\ell)=-1$ is equivalent to $g(\ell+1)=-g(\ell)$, thus we have,
\begin{equation}
R_{\ell+1}=-\frac{2\ell+5}{2\ell+3}R_{\ell}.
\end{equation}
One choice that satisfies this relation is $g(\ell)=(-1)^{\ell}\hbar \omega$. Substituting this into Eq. (8) gives,
\begin{equation}
\left \{\begin{array}{ll}
V_{1}(r,\ell)=\frac{1}{2}m\omega^{2}r^{2}+\frac{\hbar^{2}}{2m}\frac{\ell(\ell+1)}{r^{2}}+(-1)^{\ell}\hbar \omega\left(\ell+3/2\right)\\
V_{2}(r,\ell)=\frac{1}{2}m\omega^{2}r^{2}+\frac{\hbar^{2}}{2m}\frac{(\ell+1)(\ell+2)}{r^{2}}+(-1)^{\ell}\hbar \omega\left(\ell+1/2\right),
\end{array}
\right.
\end{equation}
where the CP is given by,
\begin{equation}
V(r)=\frac{1}{2}m\omega^{2}r^{2},
\end{equation}
This CP corresponds to the 3-DHO potential, but with different signs of the remainder for odd and even angular momenta. The positive remainder corresponds to odd $\ell$ and, consequently, positive eigenenergies. Conversely, the negative remainder corresponds to even $\ell$ and, therefore, negative eigenenergies. The spectrum of 3-DHO potential is rigorously known to be entirely positive and real, but the negative eigenvalues of energy for the positive 3-DHO potential may be interpreted as upside-down potentials \cite{Bender2012}.

%******************************** Subsection 3.3 : G(\ell)=\frac{\ell+1}{\ell+2} ************************************

\subsection{$G(\ell)=\frac{\ell+1}{\ell+2}$}
In this scenario, the remainder (Eq. 11) is simplified to,
\begin{equation}
R^{(\mu)}_{\ell}=\frac{\hbar^{2}(2\ell+3)}{2m(\ell+2)}\left\{\frac{w_{\mu}^{2}(r,\ell)}{\ell+2}+w_{\mu}^{\prime}(r,\ell)\right\}.
\end{equation}
The solution of this first-order differential equation for $w_{\mu}(r,\ell)=g(\ell)f_{\mu}(r)$ is,
\begin{equation}
g(\ell)=k_{0\ell}(\ell+2),
\end{equation}
where $k_{0\ell}$ is the wave number,
\begin{equation}
k_{0\ell}=\sqrt{\frac{2mR^{(\mu)}_{\ell}}{\hbar^{2}(2\ell+3)}}.
\end{equation}
The radial parts of CS are given by,
\begin{equation}
\left \{\begin{array}{ll}
f_{1}(r)=\textrm{tanh}\left\{k_{0\ell}(r+C)\right\}\\
f_{2}(r)=\textrm{coth}\left\{k_{0\ell}(r+C)\right\}\\
f_{3}(r)=-\textrm{tan}\left\{k_{0\ell}(r+C)\right\}\\
f_{4}(r)=\textrm{cot}\left\{k_{0\ell}(r+C)\right\}\\
f_{5}(r)=f_{1}(r)+f_{2}(r)\\
f_{6}(r)=f_{3}(r)+f_{4}(r)
\end{array}.
\right.
\end{equation}
As a result, the remainders are determined by,
\begin{equation}
\left \{\begin{array}{ll}
R^{(1)}_{\ell}=R^{(2)}_{\ell}=\frac{\hbar^{2}k^{2}_{0\ell}}{2m}(2\ell+3)\\
R^{(3)}_{\ell}=R^{(4)}_{\ell}=-R^{(1)}_{\ell}\\
R^{(5)}_{\ell}=-R^{(6)}_{\ell}=4R^{(1)}_{\ell}
\end{array}.
\right.
\end{equation}
This leads to the following superpartners,
\begin{equation}
\left \{\begin{array}{ll}
V^{(\mu)}_{1}(r,\ell)=\frac{\hbar^{2}k_{0\ell}}{2m}(\ell+3)g(\ell)p_{\mu}(r)-\frac{\hbar^{2}}{m}(\ell+1)\frac{w_{\mu}(r,\ell)}{r}
+\frac{\hbar^{2}}{2m}\frac{\ell(\ell+1)}{r^{2}}+s_{\mu}\frac{\hbar^{2}}{2m}g^{2}(\ell)\\
V^{(\mu)}_{2}(r,\ell)=\frac{\hbar^{2}k_{0\ell}}{2m}(\ell+1)g(\ell)p_{\mu}(r)
-\frac{\hbar^{2}}{m}(\ell+1)\frac{w_{\mu}(r,\ell)}{r}
+\frac{\hbar^{2}}{2m}\frac{(\ell+1)(\ell+2)}{r^{2}}+s_{\mu}\frac{\hbar^{2}}{2m}g^{2}(\ell),
\end{array}
\right.
\end{equation}
where,
\begin{equation}
\left \{\begin{array}{ll}
p_{1}(r)=-\textrm{sech}^{2}\left\{k_{0\ell}(r+C)\right\}, ~~~s_{1}=1,\\
p_{2}(r)=\textrm{csch}^{2}\left\{k_{0\ell}(r+C)\right\},~~~~~s_{2}=1,\\
p_{3}(r)=\textrm{sec}^{2}\left\{k_{0\ell}(r+C)\right\},~~~~~~s_{3}=-1,\\
p_{4}(r)=\textrm{csc}^{2}\left\{k_{0\ell}(r+C)\right\},~~~~~~s_{4}=-1,\\
p_{5}(r)=p_{1}(r)+p_{2}(r),~~~~~~~~~~~~s_{5}=4,\\
p_{6}(r)=p_{3}(r)+p_{4}(r),~~~~~~~~~~~~s_{6}=-4,
\end{array}
\right.
\end{equation}
Consequently, the CP can be expressed as,
\begin{equation}
V^{(\mu)}(r,\ell)=\frac{\hbar^{2}k_{0\ell}}{2m}(\ell+3)g(\ell)p_{\mu}(r)-\frac{\hbar^{2}}{m}(\ell+1)\frac{w_{\mu}(r,\ell)}{r}.
\end{equation}
These trigonometric and hyperbolic potentials represent the well-known SIPs, including Rosen-Morse II, Eckart, Scarff I, Rosen-Morse I, P\"{o}schl-Teller II, and P\"{o}schl-Teller I potentials, extended to three dimensions. These potentials are categorized as type I due to their association with hypergeometric functions \cite{gangopadhyaya1994,da2024renormalization}. Canonical transformations within this category are interconnected, and they can be transformed into one another through point canonical transformations. Here, we classify these types under the designation of "central P\"{o}schl-Teller" (CPT). According to the discussion at the end of Sec. IV, the CPT potentials are reduced to the their one dimension versions for $D$=1 and $\ell=0$. The Fig. 1 demonstrates CPT potentials $V^{(\mu)}_{1}(r,\ell)$ ($\mu$=1,2,...,6) for $\ell=1$, $\hbar^{2}=2m=1$, $k_{0\ell}=\pi$ and $C=0$. The observed behavior of the potentials, especially $V^{(1)}_{1}(r,\ell)$, corresponds to finite-range local potentials with a repulsive core, reminiscent of molecular or nuclear systems \cite{Onate2021,Koohrokhi2022}.

%******************************************Subsection 3.4 : $w(r,\ell)$ independent of $r$ ************************************

\subsection{The $r$-independent CS}
If the CS to be independent of $r$, i.e. $w(r,\ell)\equiv w(\ell)$, from Eq. (9) we have,
\begin{equation}
R_{\ell}=\frac{\hbar^{2}}{{2m}}\left\{w^{2}(\ell)-w^{2}(\ell+1)+2w(r,\ell+1)\frac{\ell+2}{r}-2w(r,\ell)\frac{\ell+1}{r}\right\}.
\end{equation}
In this case, $R_{\ell}$ is a constant only when,
\begin{equation}
2w(\ell+1)\frac{\ell+2}{r}-2w(\ell)\frac{\ell+1}{r}=0.
\end{equation}
As a result,
\begin{equation}
w(\ell+1)=w(\ell)\frac{\ell+1}{\ell+2}.
\end{equation}
It is equivalent to the case $C$, except that here the CS is independent of $r$. Thus, the CS is,
\begin{equation}
w(\ell)=k_{0\ell}(\ell+2),
\end{equation}
and by choosing the remainder as,
\begin{equation}
R_{\ell}=\frac{m(2\ell+3)}{2\hbar^{2}}\left[\frac{Z_{1}Z_{2}e^{2}}{4\pi\varepsilon_{0}(\ell+1)(\ell+2)}\right]^{2},
\end{equation}
we have the superpartners as,
\begin{equation}
\left \{\begin{array}{ll}
V_{1}(r,\ell)=-\frac{Z_{1}Z_{2}e^{2}}{4\pi\varepsilon_{0}}\frac{1}{r}
+\frac{\hbar^{2}}{2m}\frac{\ell(\ell+1)}{r^{2}}+\frac{m}{2\hbar^{2}}\left[\frac{Z_{1}Z_{2}e^{2}}{4\pi\varepsilon_{0}(\ell+1)}\right]^{2}\\
V_{2}(r,\ell)=-\frac{Z_{1}Z_{2}e^{2}}{4\pi\varepsilon_{0}}\frac{1}{r}
+\frac{\hbar^{2}}{2m}\frac{(\ell+1)(\ell+2)}{r^{2}}+\frac{m}{2\hbar^{2}}\left[\frac{Z_{1}Z_{2}e^{2}}{4\pi\varepsilon_{0}(\ell+1)}\right]^{2},
\end{array}
\right.
\end{equation}
where the CP,
\begin{equation}
V(r,\ell)=-\frac{Z_{1}Z_{2}e^{2}}{4\pi\varepsilon_{0}}\frac{1}{r},
\end{equation}
is the attractive Coulomb potential.

%***************************************** Section 4 : Broken and Unbroken SUSY ***********************************

\section{Broken and Unbroken SUSY}
By replacing the CS (Eq. 12) in the superpotential (Eq. 7), the general form of the superpotential is obtained in terms of Bessel functions. Based on Eqs. (12) and (13), if remainder $R_\ell=0$, then $B_{\ell}=0$, hence $\omega(r,\ell)=0$. In a such case, superpartners are complete isospectral and thus the SUSY is broken spontaneously. In this situation, the wave function corresponds to a spherical wave that describes a free particle in 3$D$-spherical space.

Special cases of the CS are discussed in the past section for different values of $G(\ell)$. Now, for $-1<G(\ell)<1$, we have,

$\bullet$ If $R_{\ell}>0$, then $B_{\ell}$ is imaginary, hence $\omega(r,\ell)$ is imaginary.

$\bullet$ If $R_{\ell}<0$, then $\omega(r,\ell)$ is real, but has opposite sign with $B_{\ell}$.

$\bullet$ If $\ell>-3/2$, then $A_{\ell}$ is negative.

$\bullet$ If $\ell<-3/2$, then $A_{\ell}$ is positive.

In contrast, for $G(\ell)<-1$, all the above states are reversed.

Now, we investigate conditions under which SUSY can be broken or unbroken. The radial part of the GS wave function obtain by Eq. (5), as,
\begin{equation}
u_{0\ell}(r)=N\exp\left\{-\tilde{W}(r,\ell)\right\},
\end{equation}
where $N$ is normalization constant and, the $\tilde{W}(r,\ell)$ is integral of the superpotential,
\begin{equation}
\tilde{W}(r,\ell)=\tilde{w}(r,\ell)+\tilde{w}_{\rm CefS}(r,\ell),
\end{equation}
with,
\begin{equation}
 \tilde{w}(r,\ell)=\int w(r,\ell)dr~~~~~and~~~~~\tilde{w}_{\rm CefS}(r,\ell)=-(\ell+1)\ln(r),
\end{equation}
is the integration of the CS. For a bound state, normalizability of $u_{0\ell}(r)$ requires that the wave function vanishes as $r\rightarrow a,b$, where $a$ and $b$ are the boundary points of the range $[a,b]$, respectively. Therefore, the wave function should satisfy boundary conditions $u_{0\ell}(a)=u_{0\ell}(b)=0$ and hence $\tilde{W}(r\rightarrow a,\ell)=\tilde{W}(r\rightarrow b,\ell)=\infty$. For the finite range, the CefS integration term $\tilde{w}_{\rm CefS}(r,\ell)$ is unable to provide these conditions. This implies that $\tilde{w}(r,\ell)$ should diverge to $\infty$ at the boundary points. On the other hand, for the whole range of $r$ in spherical coordinate, i.e., $[0,\infty]$, integral of CefS has following asymptotic behaviours,
\begin{equation}
\left \{\begin{array}{ll}
\textrm{for}~\ell>-1\rightarrow \left \{\begin{array}{ll}\lim\limits_{r\rightarrow\infty}\tilde{w}_{\rm CefS}(r,\ell)=-\infty\\
\lim\limits_{r\rightarrow0}\tilde{w}_{\rm CefS}(r,\ell)=\infty
\end{array}
\right.\\
\\
\textrm{for}~\ell=-1\rightarrow\tilde{w}_{\rm CefS}(r,\ell)=0\\
\\
\textrm{for}~\ell<-1 \rightarrow \left \{\begin{array}{ll}\lim\limits_{r\rightarrow\infty}\tilde{w}_{\rm CefS}(r,\ell)=\infty\\
\lim\limits_{r\rightarrow0}\tilde{w}_{\rm CefS}(r,\ell)=-\infty
\end{array}
\right.
\end{array}.
\right.
\end{equation}
Consequently, when $\ell<-1$, in the limit as $r\rightarrow0$, if $\tilde{w}(r,\ell)$ is finite or less divergent than the function $\tilde{w}_{\rm CefS}(r,\ell)$, then $\tilde{W}(r,\ell)$ approaches $-\infty$. In this scenario, which is evident in special cases A and D, SUSY is broken. Notably, in case B, the behavior is distinct because $\tilde{W}(r,\ell)\rightarrow-\infty$ as $r\rightarrow\infty$ for $\ell>-3/2$ and $R_{\ell}>0$. Consequently, with positive remainders, it is impossible to establish an unbroken SUSY system. As indicated by Eq. (19), the sign of $R_{\ell}$ changes, resulting in a negative remainder only for $\ell=2k$, where $k=0,1,2,...$. Consequently, in this case, an unbroken SUSY system is achieved solely with $R_{\ell}<0$ and even $\ell$. Thus, this case represents an upside-down version of a 3-DHO with a negative energy spectrum.

For the CPT (case $C$), the integrals of the superpotentials are as follows:
\begin{equation}
\left \{\begin{array}{ll}
\tilde{W_{1}}(r,\ell)=(\ell+2)\ln\left[\cosh\left\{k_{0\ell}(r+C)\right\}\right]-(\ell+1)\ln(r),\\
\tilde{W_{2}}(r,\ell)=(\ell+2)\ln\left[\sinh\left\{k_{0\ell}(r+C)\right\}\right]-(\ell+1)\ln(r),\\
\tilde{W_{3}}(r,\ell)=(\ell+2)\ln\left[\cos\left\{k_{0\ell}(r+C)\right\}\right]-(\ell+1)\ln(r),\\
\tilde{W_{4}}(r,\ell)=(\ell+2)\ln\left[\sin\left\{k_{0\ell}(r+C)\right\}\right]-(\ell+1)\ln(r),\\
\tilde{W_{5}}(r,\ell)=(\ell+2)\ln\left[\sinh\left\{2k_{0\ell}(r+C)\right\}\right]-(\ell+1)\ln(r),\\
\tilde{W_{6}}(r,\ell)=(\ell+2)\ln\left[\sin\left\{2k_{0\ell}(r+C)\right\}\right]-(\ell+1)\ln(r),
\end{array}
\right.
\end{equation}
The information in Fig. 2 is based on typical parameter values, namely $\ell=1$, $\hbar^{2}=2m=1$, $k_{0\ell}=\pi$, and $C=0$. According to the details provided and the perspective offered by this figure, it is evident that only $\tilde{W_{1}}(r,\ell)$ maintains the unbroken state of SUSY for these values of the parameters. The integration of CSs, in conjunction with Eq. (37) and $\Re^{(\mu)}_{0,\ell}(r)=r^{-1}u^{(\mu)}_{0,\ell}(r)$, yields the following radial wave functions,
\begin{equation}
\left \{\begin{array}{ll}
\Re^{(1)}_{0,\ell}(r)=Nr^{\ell}\left[\cosh\left\{k_{0\ell}(r+C)\right\}\right]^{-(\ell+2)},\\
\Re^{(2)}_{0,\ell}(r)=Nr^{\ell}\left[\sinh\left\{k_{0\ell}(r+C)\right\}\right]^{-(\ell+2)},\\
\Re^{(3)}_{0,\ell}(r)=Nr^{\ell}\left[\cos\left\{k_{0\ell}(r+C)\right\}\right]^{-(\ell+2)},\\
\Re^{(4)}_{0,\ell}(r)=Nr^{\ell}\left[\sin\left\{k_{0\ell}(r+C)\right\}\right]^{-(\ell+2)},\\
\Re^{(5)}_{0,\ell}(r)=Nr^{\ell}\left[\sinh\left\{2k_{0\ell}(r+C)\right\}\right]^{-(\ell+2)},\\
\Re^{(6)}_{0,\ell}(r)=Nr^{\ell}\left[\sin\left\{2k_{0\ell}(r+C)\right\}\right]^{-(\ell+2)},
\end{array}
\right.
\end{equation}
Figure 3 illustrates the radial wave functions. A 20-times magnification is applied to $\Re^{(1)}_{0,\ell}(r)$ for better visibility. As depicted in the figure, only $\Re^{(1)}_{0,\ell}(r)$ is found to be normalizable. The range of radial coordinate, angular momentum and $C$ constant for which CPT potentials are unbroken are listed in Table 1.

%************ Section 5 : D-dimensions ***********************************

\section{generalize to $D$-Dimensions}

Up to this point, we have shown that the superpotential $W(r,\ell)$ can generate various CPs, including the 3-DHO, the upside-down 3-DHO, the CPT, and the Coulomb potentials, each under different conditions. These potentials are solvable, allowing for the determination of eigenvalues of energy and eigenfunctions through analytical methods. While the current calculations are carried out in 3D spherical coordinates, they can be readily extended to arbitrary D-dimensions by implementing the following replacement for angular momentum \cite{Dong2007},
\begin{equation}
\ell\rightarrow\ell+\frac{D-3}{2}.
\end{equation}
With this choice, the superpotential is generalized to,
\begin{equation}
W(r,\ell,D)=\frac{\hbar}{\sqrt{2m}}\left\{w(r,\ell,D)-\frac{\ell+\frac{D-1}{2}}{r}\right\}.
\end{equation}
For example, for 3-DHO, by replacing $\ell^{'}=\ell-C$, and $C=-\frac{D-3}{2}$ the GS wave function,
\begin{equation}
u_{0\ell^{'}}=Nr^{\ell^{'}+\frac{D-1}{2}}\exp\left\{-\frac{m\omega}{2\hbar}r^{2}\right\}
\end{equation}
diverges for $\ell^{'} < \frac{D-1}{2}$ at the origin $r=0$, and hence corresponds to the case of broken SUSY.

The CPT potentials can now be generalized to any $D$-dimension by substituting Eq. (44) into Eq. (28), namely,
\begin{equation}
V^{(\mu)}(r,\ell,D)=\frac{\hbar^{2}k_{0\ell}}{2m}\left(\ell+\frac{D+3}{2}\right)g(\ell,D)p_{\mu}(r)
-\frac{\hbar^{2}}{m}\left(\ell+\frac{D-1}{2}\right)\frac{w_{\mu}(r,\ell,D)}{r}.
\end{equation}
where,
\begin{equation}
w_{\mu}(r,\ell,D)=g(\ell,D)f_{\mu}(r)~~~ ,~~~ g(\ell,D)=k_{0\ell}\left(\ell+\frac{D+1}{2}\right)
\end{equation}
It is important to note that the only choice for angular momentum in one dimension, $D=1$, is zero $\ell=0$. In the such condition the Cefs is zero and the superpotential, Eq. (44), equals with CS (except a coefficient). As a consequence, Eq. (46) for $\mu=1$ is reduced to the one-dimensional version,
\begin{equation}
V^{1}(r)=-\frac{\hbar^{2}k^{2}}{m}\textrm{sech}^{2}\{k(r+C)\},
\end{equation}
here the wave number is $k\equiv k_{00}$.
%******************************************Section 5 : Hierarchy of the CPT Potential ************************************

\section{Hierarchy of the CPT Potential}
In SUSY-QM, superpartners, $V_{1}(r,\ell)$ and $V_{2}(r,\ell)$ are related to a superpotential $W(r,\ell)$ via RE, which now we affix it an subscript "1" as $W_{1}(r,\ell)$. The RE which is derived by Hamiltonian factorization method can be generalized to obtain a hierarchy of Hamiltonians $H_{1}, H_{2}, ... , H_{n+1}$, ($n$=0,1,2,...) so that each Hamiltonian has one fewer bound state than before $E^{(n+1)}_{m+1}=E^{(n+2)}_{m}$, ($m$=0,1,2,...). As a result, if one has an exact solution to the potential problem for $H_{1}$, one can find the wave functions and eigenvalues for the entire hierarchy of Hamiltonians. In this way, $n$th potential $V_{n}(r,\ell)$ and superpotential $W_{n}(r,\ell)$ are related as follows,
\begin{equation}
V_{n+2}(r,\ell)=W^{2}_{n+2}(r,\ell)-\frac{\hbar}{\sqrt{2m}}W^{\prime}_{n+2}(r,\ell)+E^{(n+2)}_{0,\ell}
=W^{2}_{n+1}(r,\ell)+\frac{\hbar}{\sqrt{2m}}W^{\prime}_{n+1}(r,\ell)+E^{(n+1)}_{0,\ell},
\end{equation}
where $E^{(n)}_{0,\ell}$ is the GSE of $n$th Hamiltonian $H_{n}$. Recall zero GSE of $H_{1}$, i.e. $E^{(1)}_{0,\ell}=0$ is a necessary condition for unbroken SUSY. For the $n$th CPT superpotential ($\mu=1$), we obtain,
\begin{equation}
w_{n+1}(r,\ell)=k_{n,\ell}(\ell+n+2)\textrm{tanh}\left\{k_{n,\ell}(r+C)\right\},
\end{equation}
where the condition,
\begin{equation}
g(\ell+n+1)=G_{n+1}(\ell)g(\ell+n),
\end{equation}
with,
\begin{equation}
G_{n+1}(\ell)=\frac{\ell+n+1}{\ell+n+2},
\end{equation}
requires the following relation for consecutive wave numbers,
\begin{equation}
k_{n+1,\ell}(\ell+n+3)=k_{n,\ell}(\ell+n+1).
\end{equation}
According our calculations, $n$th remainder is attained by,
\begin{equation}
R_{n+1,\ell}=E_{0,\ell}^{(n+2)}-E_{0,\ell}^{(n+1)}=\left\{2(\ell+n+2)+1\right\}k^{2}_{n,\ell}-2\left(k^{2}_{n-1,\ell}+k^{2}_{n-2,\ell}+...+k^{2}_{0\ell}\right).
\end{equation}
As a result, $m$th energy level of $n$th Hamiltonian is achieved as,
\begin{equation}
E^{(n+1)}_{m+1,\ell}=E^{(n+2)}_{m,\ell}=\sum_{i=m}^{n}R_{i+1,\ell}.
\end{equation}
Ultimately, $n$th CPT potential is,
\begin{equation}
\begin{array}{ll}
V_{n+1}(r,\ell)=-\frac{\hbar^{2}k^{2}_{n,\ell}}{2m}(\ell+n+2)(\ell+n+3)\textrm{sech}^{2}\left\{k_{0\ell}(r+C)\right\}
-2\frac{\hbar^{2}k_{n,\ell}}{2m}(\ell+n+1)(\ell+n+2)\frac{\textrm{tanh}\left\{k_{0\ell}(r+C)\right\}}{r}\\
~~~~~~~~+\frac{\hbar^{2}}{2m}\frac{(\ell+n)(\ell+n+1)}{r^{2}}-E_{0,\ell}^{(n+1)}.
\end{array}
\end{equation}

%******************************************Section 6 : Complex Non-Hermitian Superpotential ************************************

\section{Complex Isospectral Deformations}
Recently, a formalism has been introduced through which SIPs can be extended to the complex domain \cite{Kooh2023}. Now, we utilize this formalism to derive the complex non-Hermitian versions of the superpotential and subsequently superpartners. To initiate this process, we introduce an imaginary term $i w_{i}(r)$ to the superpotential,
\begin{equation}
W(r,\ell)=\frac{\hbar}{\sqrt{2m}}\left\{w(r,\ell)-\frac{\ell+1}{r}+iw_{i}(r)\right\},
\end{equation}
As a consequence, the superpartners also become complex, causing their real parts to undergo a change to,
\begin{equation}
\left \{\begin{array}{ll}
V_{1r}(r,\ell)=\frac{\hbar^{2}}{{2m}}\left\{w^{2}(r,\ell)-w_{i}^{2}(r)-w^{\prime}(r,\ell)-2w(r,\ell)\frac{\ell+1}{r}+\frac{\ell\left(\ell+1\right)}{r^{2}}\right\}\\
V_{2r}(r,\ell)=\frac{\hbar^{2}}{{2m}}\left\{w^{2}(r,\ell)-w_{i}^{2}(r)+w^{\prime}(r,\ell)-2w(r,\ell)\frac{\ell+1}{r}+\frac{\left(\ell+1\right)\left(\ell+2\right)}{r^{2}}\right\}.
\end{array}
\right.
\end{equation}
and their imaginary components are determined by,
\begin{equation}
\left \{\begin{array}{ll}
V_{1i}(r,\ell)=\frac{\hbar^{2}}{{2m}}\left\{2w_{i}(r)\left[w(r,\ell)-\frac{\ell+1}{r}\right]-w_{i}^{\prime}(r)\right\}\\
V_{2i}(r,\ell)=\frac{\hbar^{2}}{{2m}}\left\{2w_{i}(r)\left[w(r,\ell)-\frac{\ell+1}{r}\right]+w_{i}^{\prime}(r)\right\}.
\end{array}
\right.
\end{equation}
By comparing Eqs. (8) and (59), we recognize that the real term of the CP is generalized to,
\begin{equation}
V_{r}(r,\ell)=V(r,\ell)-\frac{\hbar^2}{2m}w^{2}_{i}(r),
\end{equation}
Therefore, as evident from the equations, the primary challenge lies in obtaining the imaginary part of the CS. The remainder plays a crucial role in determining it. According to Eq. 9, as a consequence of complexifying the superpartners, it transforms into a complex quantity as,
\begin{equation}
R=V_{2}(r,\ell)-V_{1}(r,\ell+1)=R_{\ell}+iR_{i}(r),
\end{equation}
where $R_{\ell}$ represents the real part of the remainder (Eq. 11), and $R_{i}(r)$ denotes its imaginary component, as follows,
\begin{equation}
R_{i}(r)=\frac{\hbar^{2}}{m}\left(\left[1-G(\ell)\right]w(r,\ell)+\frac{1}{r}\right)w_{i}(r)+w^{\prime}_{i}(r),
\end{equation}
By enforcing the nullification of the remainder imaginary part, $R_{i}(r)=0$, we achieve two objectives: first, we restore the condition of shape invariance to the superpartners, and second, we ensure that their energy spectra are real. With this condition, we can determine the imaginary part of the CS as,
\begin{equation}
w_{i}(r)=\frac{C_{i}}{r}\exp\left\{[G(\ell)-1]\tilde{w}(r,\ell)\right\},
\end{equation}
Similarly, the imaginary component of CP is derived from Eq. (59). The CS and CP components of the special cases explored in Sec. III are outlined in Table 2. It is noteworthy that complex CPs share isospectrality with real CPs.

%******************************************Section 6 : Deuteron ************************************

\section{An Application: Deuteron}
Deuteron is a two-nucleon bound system. Based on their separation distances, a two-nucleon interaction usually is divided into three regions: the short-range, the medium-range, and the long-range regions. In the short-range region, due to the incompressibility of nuclear matter, the two nucleons wave functions cannot completely overlap  \cite{WANG2018207}. This part of the interaction is described by a potential that has a repulsive term in its core $\lim\limits_{r\rightarrow 0} V(r)\rightarrow \infty$. On the other hand, since the force between two nucleons is a finite range force, the interaction potential vanishes at long distances $\lim\limits_{r\rightarrow\infty} V(r)\rightarrow 0$. However, existence of an attractive well between the these short and long regions, i.e. the medium-range region, is essential to form a bound system. Fortunately, CPT potential satisfies all the mentioned regions by a smooth function (see Fig. 1).

For most practical applications, the deuteron is typically described by an effective potential due to the convoluted nature of nucleon interactions. It has also been proved that the GS of the deuteron consists of two states $^{13}S_{1}$ and $^{13}D_{1}$, which means nuclear force combines the two states with two different angular momenta, namely, $\ell=0$ and $\ell=2$. In previous research, we obtained an effective angular momentum $\ell_{eff}$ \cite{Koohrokhi2022}. According to Eq. (56), we introduce an effective potential for deuteron's GS $n=0$, with an effective angular momentum $\ell \equiv \ell_{eff}$, and an effective wave number $k_{0\ell} \equiv k_{eff}$, as follows,
\begin{equation}
\begin{array}{ll}
V_{eff}(r)=-\frac{\hbar^{2}k^{2}_{eff}}{2m}(\ell_{eff}+2)(\ell_{eff}+3)\textrm{sech}^{2}\left\{k_{eff}(r+C)\right\}
-2\frac{\hbar^{2}k_{eff}}{2m}(\ell_{eff}+1)(\ell_{eff}+2)\frac{\textrm{tanh}\left\{k_{eff}(r+C)\right\}}{r}\\
~~~~~~~~+\frac{\hbar^{2}}{2m}\frac{\ell_{eff}(\ell_{eff}+1)}{r^{2}}-E_{0},
\end{array}
\end{equation}
where $m$ is the reduced mass of the proton-neutron system. We get the effective wave number $k_{eff}$ by equating the constant terms of the two Eqs. (27) and (64), as,
\begin{equation}
E_{0}=-\frac{\hbar^{2}k^{2}_{eff}}{2m}(\ell_{eff}+2)^{2}.
\end{equation}
To obtain the coefficient $C$, we use the assumption that the wave function is maximum at $r=R_{m}$, i.e., $\Re^{\prime}(R_{m})=0$. This condition in terms of Eqs. (50) and (7), yields,
\begin{equation}
k_{eff}(\ell_{eff}+2)\textrm{tanh}\left\{k_{eff}(R_{m}+C)\right\}-\frac{\ell_{eff}}{R_{m}}=0,
\end{equation}
Ultimately, the radial wave function of deuteron is obtained from Eq. (42), for $\mu=1$, $\ell\equiv \ell_{eff}$, and $k_{0\ell}\equiv k_{eff}$,
\begin{equation}
\Re(r)=Nr^{\ell_{eff}}\left[\cosh\left\{k_{eff}(r+C)\right\}\right]^{-(\ell_{eff}+2)},
\end{equation}
where normalizaion constant $N$ is obtained by $\int_{0}^{\infty}\Re^{2}(r)r^{2}\textrm{d}r=1$. The taken values of $R_{m}$, $E_{0}$, and $\ell_{eff}$ \cite{Koohrokhi2022}, as well as the calculated values $k_{eff}$, $m$, $C$, and $N$ are listed in Table 3. The Fig. 4 illustrates the deuteron's effective potential $V_{eff}(r)$ (Eq. 64) and the wave function $\Re(r)$ (Eq. 67) 200-fold magnified.

%******************************************Section 7 : Conclusion ************************************

\section{Conclusion}
The unified scheme not only sets a family of solvable CPs but also reveals some facts of the origin of interactions. A homogeneous isotropic CS produces the Coulomb potential whereas an isotropic local CS originates an effective nuclear potential. However, both Coulomb and nuclear CSs have the same $\ell$ dependance, i.e. $w(\ell)=k_{0\ell}(\ell+2)$. Certainly, the nuclear force and consequently the two-nucleon potential are not central. In addition, a nuclear potential includes other significant aspects, e.g. the tensor term. Nevertheless, a complicated nuclear potential can replace with an effective potential as a proper approximation and reflect the static properties of a bound two-nucleon system, i.e. deuteron  \cite{Koohrokhi2022}.

In the realm of non-Hermitian parity-time ($\mathcal{PT}$)-symmetric potentials, upside-down potentials stand out as a notable consequence \cite{bender2018}. In one dimension, a widely recognized general form is given by $V(x) = x^{2}(ix)^{\epsilon}$, where $\epsilon$ is a real-positive parameter. Notably, for $\epsilon=1$, this yields the imaginary cubic potential $V(x) = ix^{3}$, and for $\epsilon=2$, the resulting real non-Hermitian quartic potential is $V(x) = -x^{4}$. The eigenvalues of these potential are noteworthy, as they are all real, positive, and discrete. The latter potential exemplifies the upside-down potential type, expressed as $V(x) = -|x|^{2p}$, where $p\geq 2$ is an integer. Recently, these upside-down potentials, beyond their theoretical significance, have garnered experimental attention, particularly in the context of $\mathcal{PT}$ phase transition in reflectionless quantum scattering \cite{soley2023}. Conversely, as demonstrated in our findings, a positive potential manifests a real, negative, and discrete spectrum in dimensions higher than one dimension. This observation holds significance both theoretically and in terms of potential experimental applications, shaping the trajectory of future studies in this domain.

As discussed, upside-down potentials emerge as a consequence of generalizing real potentials to the complex domain. This extension is not solely a mathematical process; it is also expanding the horizons of our understanding of real quantum systems. For instance, as detailed in Sec. VII, during the extension to complex space, not only is an imaginary term introduced to both CS and CP, but also the real component of CP undergoes a modification. Hence, these novel manifestations of CPs, such as upside-down potentials, may exhibit seemingly unconventional outcomes, the rationale for which is rooted in the extension to the complex domain.

The predominant focus of achievements and research, whether related to SIPs or $\mathcal{PT}$-symmetric potentials, has largely centered around one-dimensional scenarios. Nevertheless, real physical conditions are not always confined to the one dimension, making the exploration of quantum systems in higher dimensions essential. In the context of central symmetry, as investigated towards the conclusion of Sec. V, the impact of adding each spatial dimension is analogous to augmenting one-half of the angular momentum unit. Specifically, the influence of even dimensions, $D=2n~(n=1,2...)$, is comparable to that of half-integer angular momenta, while the impact of odd dimensions, $D=2n+1$, is akin to that of integer angular momenta.

%*************************Table 1*************************
\begin{table}[tbp]
\caption{The ranges in which the CPT potentials are unbroken.}
\begin{center}
\begin{tabular}{ |c|c|c|c|c|c|c|c|c|c|c|c| }
\hline \scriptsize{CPT} & \scriptsize{$r$} &
\scriptsize{$\ell$}  & \scriptsize{$C$}\\
\hline \scriptsize{$\tilde{W_{1}}$} &
\scriptsize{[0,$\infty$]} & \scriptsize{$\ell>-1$}  & \scriptsize{real} \\
\scriptsize{ } & \scriptsize{ } &
\scriptsize{ }  & \scriptsize{ }\\
\scriptsize{$\tilde{W_{2}}$} &
\scriptsize{[0,$\infty$]} & \scriptsize{$\ell>-1$}  & \scriptsize{$C>0$} \\
\scriptsize{ } & \scriptsize{ } &
\scriptsize{ }  & \scriptsize{ }\\
\scriptsize{$\tilde{W_{3}}$} &
\scriptsize{$\left[\mid\frac{n\pi}{k_{0\ell}}-C\mid,\mid\frac{(n+1)\pi}{k_{0\ell}}-C\mid\right]$} & \scriptsize{$\ell<-2$}  & \scriptsize{real} \\
\scriptsize{ } & \scriptsize{ } &
\scriptsize{ }  & \scriptsize{ }\\
\scriptsize{$\tilde{W_{4}}$} &
\scriptsize{$\left[\mid\frac{(2n+1)\pi}{k_{0\ell}}-C\mid,\mid\frac{(2n+3)\pi}{k_{0\ell}}-C\mid\right]$} & \scriptsize{$\ell<-2$}  & \scriptsize{real} \\
\scriptsize{ } & \scriptsize{ } &
\scriptsize{ }  & \scriptsize{ }\\
\scriptsize{$\tilde{W_{5}}$} &
\scriptsize{[0,$\infty$]} & \scriptsize{$\ell>-1$}  & \scriptsize{$C>0$} \\
\scriptsize{ } & \scriptsize{ } &
\scriptsize{ }  & \scriptsize{ }\\
\scriptsize{$\tilde{W_{6}}$} &
\scriptsize{$\left[\mid\frac{(2n+1)\pi}{k_{0\ell}}-C\mid,\mid\frac{(2n+3)\pi}{k_{0\ell}}-C\mid\right]$} & \scriptsize{$\ell<-2$}  & \scriptsize{real} \\
\hline
\end{tabular}
\end{center}
\end{table}

%*************************Table 2*************************
\begin{table}[tbp]
\caption{The imaginary terms of CSs and CPs. Here $n$ is an integer number.}
\begin{center}
\begin{tabular}{ |c|c|c|c|c|c|c|c|c|c|c|c| }
\hline \scriptsize{CP} & \scriptsize{$G(\ell)$} &
\scriptsize{$w_{i}(r)$}  & \scriptsize{$V_{i}(r,\ell)$}\\
\hline \scriptsize{3-DHO} &
\scriptsize{1} & \scriptsize{$\frac{C_{i}}{r}$}  & \scriptsize{$C_{i}\left\{\hbar \omega-\frac{\hbar^{2}}{2m}\left(\frac{2\ell+3-2C}{r^{2}}\right)\right\} $} \\
\scriptsize{ } & \scriptsize{ } &
\scriptsize{ }  & \scriptsize{ }\\
\scriptsize{upside-down 3-DHO} &
\scriptsize{-1} & \scriptsize{$\frac{C_{i}}{r}\exp\left\{-\frac{m\omega}{\hbar}r^{2}\right\}$} & \scriptsize{$ -C_{i}\frac{\hbar^{2}}{2m}(2\ell+1)\frac{\exp\left\{-\frac{m\omega}{\hbar}r^{2}\right\}}{r^2}$} \\
\scriptsize{ } & \scriptsize{ } &
\scriptsize{ }  & \scriptsize{ }\\
\scriptsize{Coulomb} &
\scriptsize{$\frac{\ell+1}{\ell+2}$} & \scriptsize{$\frac{C_{i}}{r}\exp(-k_{0\ell}r)$}  & \scriptsize{$C_{i}\frac{\hbar^{2}}{2m}(2\ell+3)\left\{k-\frac{1}{r}\right\}\frac{\exp(-k_{0\ell}r)}{r}$}\\
\scriptsize{ } & \scriptsize{ } &
\scriptsize{ }  & \scriptsize{ }\\
\scriptsize{CPT} &
\scriptsize{$\frac{\ell+1}{\ell+2}$} & \scriptsize{$\left \{\begin{array}{ll}\frac{C_{i}}{r}\textrm{sech}\left\{k_{0\ell}(r+C)\right\}\\
\frac{C_{i}}{r}\textrm{csch}\left\{k_{0\ell}(r+C)\right\}\\
\frac{C_{i}}{r}\sec\left\{k_{0\ell}(r+C)\right\}\\
\frac{C_{i}}{r}\csc\left\{k_{0\ell}(r+C)\right\}\\
\frac{C_{i}}{r}\textrm{csch}\left\{2k_{0\ell}(r+C)\right\}\\
\frac{C_{i}}{r}\csc\left\{2k_{0\ell}(r+C)\right\}
\end{array}
\right.$}  & \scriptsize{$ \left \{\begin{array}{ll}\frac{C_{i}\hbar^{2}\textrm{sech}\left\{k_{0\ell}(r+C)\right\}}{2mr}\left(k_{0\ell}(2\ell+5)\textrm{tanh}\left\{k_{0\ell}(r+C)\right\}-\frac{2\ell+1}{r}\right)\\
\frac{C_{i}\hbar^{2}\textrm{csch}\left\{k_{0\ell}(r+C)\right\}}{2mr}\left(k_{0\ell}(2\ell+5)\textrm{coth}\left\{k_{0\ell}(r+C)\right\}-\frac{2\ell+1}{r}\right)\\
-\frac{C_{i}\hbar^{2}\textrm{sec}\left\{k_{0\ell}(r+C)\right\}}{2mr}\left(k_{0\ell}(2\ell+5)\textrm{tan}\left\{k_{0\ell}(r+C)\right\}-\frac{2\ell+1}{r}\right)\\
\frac{C_{i}\hbar^{2}\textrm{csc}\left\{k_{0\ell}(r+C)\right\}}{2mr}\left(k_{0\ell}(2\ell+5)\textrm{cot}\left\{k_{0\ell}(r+C)\right\}-\frac{2\ell+1}{r}\right)\\
\frac{C_{i}\hbar^{2}\textrm{csch}\left\{2k_{0\ell}(r+C)\right\}}{2mr}\left(2k_{0\ell}(2\ell+5)\textrm{coth}\left\{2k_{0\ell}(r+C)\right\}-\frac{2\ell+1}{r}\right)\\
\frac{C_{i}\hbar^{2}\textrm{csc}\left\{2k_{0\ell}(r+C)\right\}}{2mr}\left(2k_{0\ell}(2\ell+5)\textrm{cot}\left\{2k_{0\ell}(r+C)\right\}-\frac{2\ell+1}{r}\right)
\end{array}
\right.$}\\
\hline
\end{tabular}
\end{center}
\end{table}

%*************************Table 3*************************
\begin{table}[tbp]
\caption{Static properties of deuteron.}
\begin{center}
\begin{tabular}{ |c|c|c|c|c|c|c|c|c|c|c|c| }
\hline \scriptsize{$E_{0}$ (MeV)} & \scriptsize{$m$ (MeV/c$^{2}$)} &
\scriptsize{$k_{eff} (fm^{-1})$} & \scriptsize{$\ell_{eff}$} & \scriptsize{$C$ (fm)}
& \scriptsize{$N$} & \scriptsize{$R_{m}$ (fm)}\\
\hline \scriptsize{-2.22456627} &
\scriptsize{469.459} & \scriptsize{0.110} & \scriptsize{0.1029} &
\scriptsize{1.489} & \scriptsize{0.0857} & \scriptsize{1.4295}\\
\hline
\end{tabular}
\end{center}
\end{table}

\clearpage
%*************************fig. 1*************************

\begin{figure}
  \includegraphics[width=0.95\textwidth]{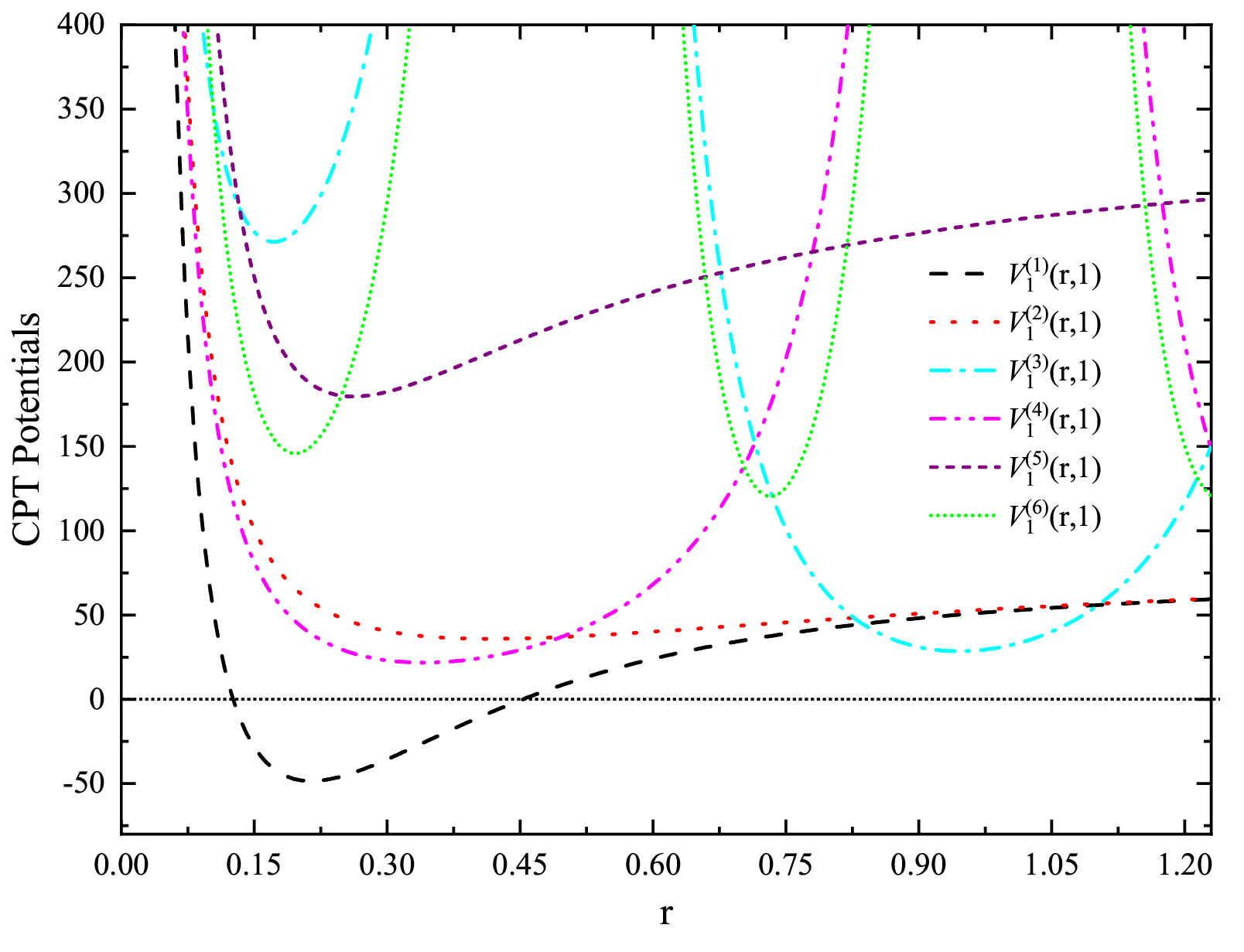}
\caption{The CPT potentials (Eq. 26) $V^{(\mu)}_{1}(r,\ell)$ ($\mu$=1,2,...,6) for $\ell=1$, $\hbar^{2}=2m=1$, $k_{0\ell}=\pi$ and $C=0$.} \label{fig:1}
\end{figure}

%*************************fig. 2*************************
\begin{figure}
  \includegraphics[width=0.95\textwidth]{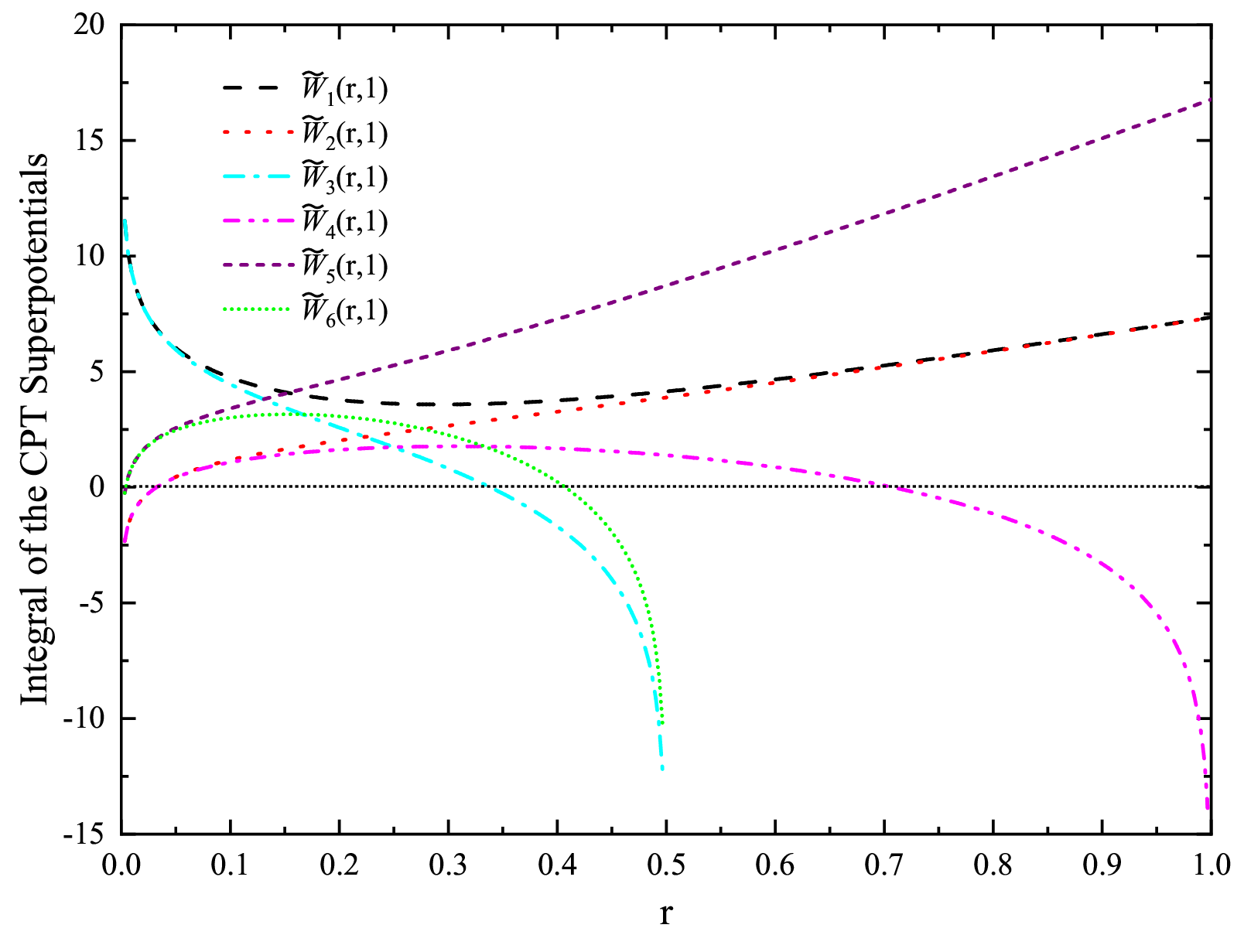}
\caption{Integral of CS (Eq. 39) $\tilde{W_{\mu}}(r,\ell)$ ($\mu$=1,2,...,6) of CPT for $\ell=1$, $\hbar^{2}=2m=1$, $k_{0\ell}=\pi$ and $C=0$.} \label{fig:2}
\end{figure}
\clearpage
%*************************fig. 3*************************
\begin{figure}
  \includegraphics[width=0.95\textwidth]{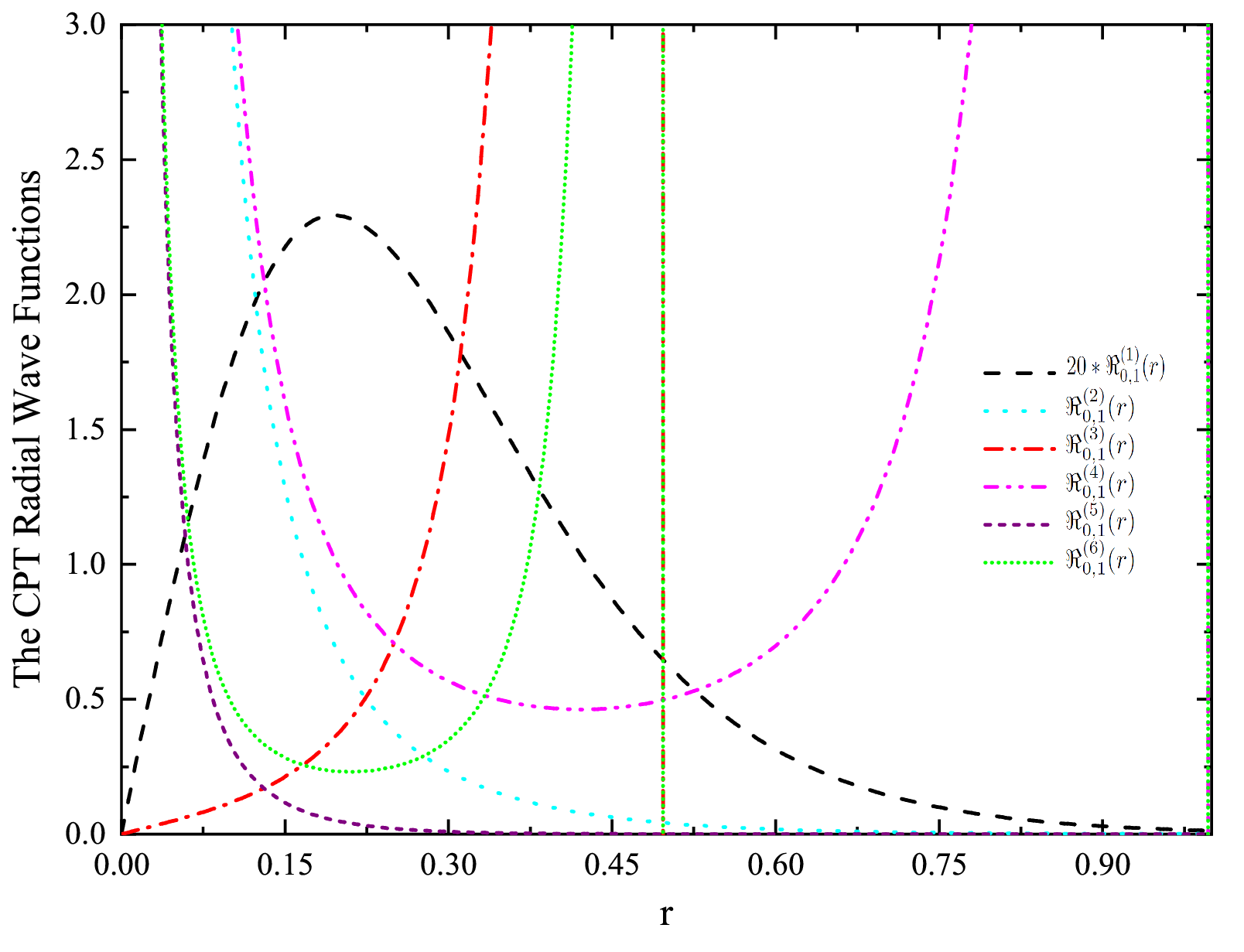}
\caption{The CPT radial wave functions (Eq. 40) $\Re^{(\mu)}_{0,\ell}(r)$ ($\mu$=1,2,...,6) for $N=1$, $\ell=1$, $\hbar^{2}=2m=1$, $k_{0\ell}=\pi$ and $C=0$.} \label{fig:3}
\end{figure}
\clearpage
%*************************fig. 4*************************

\begin{figure}
  \includegraphics[width=0.95\textwidth]{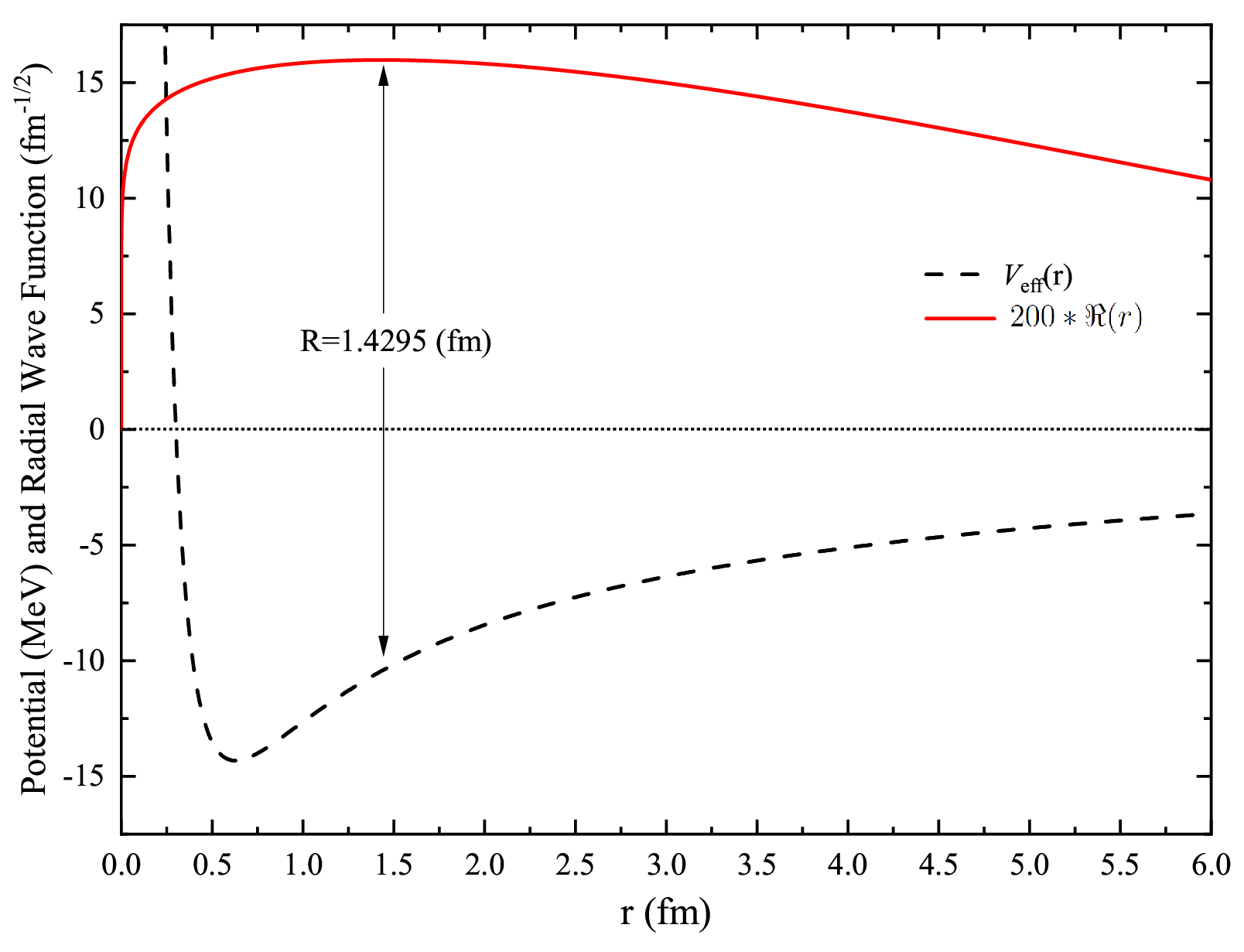}
\caption{The effective potential $V(r)$ and ground state wave function $\Re(r)$.} of deuteron\label{fig:4}
\end{figure}

%\bibliography{mybibfile}

\end{document}